# The Opening Scholarly Communication in Social Sciences project OSCOSS

Philipp Mayr[1], Christoph Lange

## Abstract

This paper outlines the main ideas and objectives of the DFG-funded project "Opening Scholarly Communication in the Social Sciences" (OSCOSS). The paper makes up with a problem statement discussing the traditional scholarly publication process and the shortcomings of the portable document format (PDF) as publication format. In the following, the main use cases of the project are introduced: readers, authors and reviewers of scientific publications. The main part of the paper is describing the project objectives and the architecture of the intended project software infrastructure. Project objectives are: 1) establishing electronic papers as the focal points of scholarly communication, 2) integrating social science research data and the paper in a closer way, 3) keep up the high layout and formatting standards and 4) provide publishers with a technical infrastructure.

## 1. Problem Statement

For over 350 years, scientific articles have been the primary means by which scholars have formally communicated their work such as hypotheses, methods, results or experiments. This includes the social sciences, where data often play a key role in articles. Advances in technology made it possible for the scientific article to adopt electronic dissemination channels. However, the increasingly collaborative scientific process that leads to the publication of an article is still insufficiently supported by contemporary electronic information systems.
The OSCOSS project[2] (Opening Scholarly Communication in Social Sciences), which will be outlined in the following, aims at providing integrated support for all steps of the scholarly communication process:
1. collaborative writing of a scientific paper,
2. collecting data related to existing publications,
3. interpreting and including data in a paper,
4. submitting the paper for peer review,
5. reviewing the paper,
6. publishing an article, and, finally,
7. facilitating its consumption by readers.

---

[1] Philipp Mayr was a student at the Berlin School of Library and Information Science and has attended a few courses of Konrad Umlauf. As a teacher, Konrad Umlauf was always an excellent supporter of open accessible information and techniques to assess the quality of single pieces of information. E.g. Konrad Umlauf published the "Berliner Handreichungen zur Bibliotheks- und Informationswissenschaft", an open access series for theses of the institute which is dating back to 1992. The author has applied some of these techniques in his doctoral dissertation (Mayr 2009).
[2] All software and test datasets in the OSCOSS project are published under https://github.com/OSCOSS.

The OSCOSS project will support this process considering in particular the perspective of three main actors detailed in the use case descriptions below: readers, authors and reviewers. Currently, these actors adopt several separate electronic information systems, which support certain single steps of their work. For example, mailing lists and project management systems facilitate the discussion of ideas and the planning of research tasks. Open tools for statistical analysis and data visualisation such as R[3] help authors to validate hypotheses and thus obtain research results. Web-based word processors such as Google Docs, Overleaf or Fidus Writer[4] facilitate collaborative writing. Electronic submission systems such as Open Journal Systems (OJS[5]) facilitate peer reviewing. Web content management systems (CMSs) enable users without any technical knowledge to create websites and to upload documents. Further, social network features of CMSs such as TYPO3 enable readers to discuss web publications with their authors, or to recommend them to their peers.

Still, media discontinuities between these steps cause inefficiency and loss of information. For example, word processors lack direct access to data that authors may want to interpret in a paper. Also, while authors have a rich and detailed understanding of the structure and characteristics of the data that a document is based on, word processors cannot currently capture such structural information, as they have been designed for layouting documents to be consumed by human readers only, not for enabling information systems to manage knowledge. Peer reviewers cannot provide feedback inside the same environment in which authors wrote their papers. Open access web publishing is often constrained to document formats that have been designed for paper printing, such as Portable Document Format (PDF), but that neglect the Web's full accessibility and interactivity potential. Finally, readers are forced to see a single frozen view of the underlying data in a paper; they are unable to access the full extent and the further dimensions of the data and to make their own observations beyond the restricted scope chosen by the author.

## 2. Use cases

To justify the relevance of our work and to explain in which ways our project opens scholarly communication and publishing from multiple perspectives, we first analyse one real world example, and then present three idealised use cases from the perspective of three major stakeholders of scholarly communication and publishing: reader, author and reviewer.

**Real world example:** In a recent article in the *mda* journal[6], Kroll (2011) cites a dataset by name, writing "Der Index wird zuletzt auf Basis der Daten der BIBB/BAuA-Erwerbstätigen-befragung 2006 und des Telefonischen Gesundheitssurveys „Gesundheit in Deutschland Aktuell" (GEDA) 2009 des Robert Koch-Instituts anhand von Gesundheitsindikatoren intern und extern validiert."
He also cites other articles citing the same dataset ("zu den Datensätzen vgl. Hall 2009; Kurth et al. 2009; RKI 2010"; Kroll 2001, p.67). But an explicit link to the dataset is missing in his article. Still, the dataset is registered in *da|ra*[7], a registry for datasets, and can be looked

---

[3] https://www.r-project.org/
[4] http://www.fiduswriter.org, a "semantic word processor for academics", which we are using and extending in OSCOSS.
[5] https://pkp.sfu.ca/ojs/
[6] methods, data, analyses (http://www.gesis.org/publikationen/zeitschriften/mda/)
[7] Registration agency for social and economic data (http://www.da-ra.de/)

up; it has a metadata record and a DOI. In the concrete case the datasets[8,9] can be analysed after signing a contract.

**UC1: Reader.** Mark is conducting a survey on "paradata"[10]. He has selected several relevant articles that have been published recently in *mda* and is now studying them in detail. He wants to focus on observations with a high statistical significance. In one article, he has identified an interesting "non-response bias", which is presented in a table. For his future work, he wants to precisely bookmark this occurrence as "useful for my survey", and add an annotation that helps him remember what exactly was interesting and why.

**UC2: Author.** Jakob has a draft of a paper, written in Word, and wants to extend it by performing a different analysis on those base data that Arthur, a researcher from the same community, has used for an earlier publication. Arthur, in his publication, cites a dataset from *da|ra*. The dataset has a DOI, and furthermore, Arthur describes in his paper what chunks of the dataset his analysis based on, and what analysis method he applied. The R code that implements the data analysis is open source and available from a source code repository hosted by GitHub. Arthur presented the output of this analysis as a table and a diagram in his paper. Jakob invites his co-author Dagmar to do a *different* analysis of the same data: Dagmar re-applies the same analysis method but changes the values of some regression parameters, and Jakob compares the result of this analysis to the result of Arthur's analysis. Their new article includes a new table, and a copy of Arthur's one, side by side and citing Arthur's original table and the underlying dataset, and draws new conclusions. They submit their article to *mda*.

**UC3: Reviewer.** Jakob and Dagmar submitted their manuscript to mda, pointing to the data and the R code. Rainer gets assigned the manuscript for review. He wants to check whether Jakob and Dagmar have done their analysis in a correct way. He downloads their R code and raw data and redoes the calculation described. He observes that, for one of the statements that Jakob and Dagmar have made in the results section of their manuscript, the R output of the analysis does not give sufficient evidence. They should have known from Ariane's paper published two years ago, that in one of the items of the dataset they analysed, some data items are too sparse for reliably applying the significance test to them. Rainer marks the respective statement in Jakob's and Dagmar's result section, adding a reference to the methodology section of Ariane's paper and to the affected item in the dataset. Finally, the editorial board decides to accept the submission, provided that a major revision is made. Jakob and Dagmar receive the paper with 100 comments attached. As the comments are attached to precise parts of the paper, grouped by reviewers and classified as "major" vs. "minor", they can quickly prioritise the necessary tasks to improve their article.

## 3. Starting Point

Web technologies, which turn data on the Web into machine-comprehensible knowledge, can address the problems pointed out initially. Isolated solutions exist already in the social sciences. There are, for example, tools for publishing data according to Web standards, to facilitate their retrieval and visualisation. However, such tools have not yet been integrated with other tools for writing, reviewing, publishing and reading articles.

---

[8] BIBB/BAuA- Erwerbstätigenbefragung 2006 http://dx.doi.org/10.7803/501.06.1.8.11
[9] GEDA 2009 http://dx.doi.org/10.7797/26-200809-1-1-2
[10] The paradata of a survey are data about the process by which the survey data were collected. Paradata are usually "administrative data about the survey." (http://en.wikipedia.org/wiki/Paradata)

Tools that **connect articles to structured data** have not existed in the social sciences so far. There are general-purpose Semantic Digital Libraries (SDLs) (Hienert et al. 2015), which aim at providing uniform access to metadata and, partly, to document contents, to better support information retrieval and classification tasks. Research prototypes such as JeromeDL (Kruk et al. 2007) allow users to annotate books, papers, and resources. Production-scale systems, such as the German Digital Library[11] (Deutsche Digitale Bibliothek), have so far focused on metadata rather than fine-grained structures *within* documents. Fine-grained connections between documents and data have so far most successfully been investigated in the life sciences, including research towards improving their interoperability and user experience. The DOMEO annotation tool (Ciccarese et al. 2012) allows users to manually or semi-automatically create unstructured or semi-structured annotations that can be private, shared within selected groups, or public. The Living Document annotation environment (Garcia-Castro et al. 2010) allows for turning a document into an interface to explore the Web, i.e. to online sources of data and knowledge related to the topic of the document. Utopia Documents (Attwood et al. 2010) is a desktop-based PDF reader with similar functionalities.

*We aim at transferring these ideas to the social sciences by integrating existing, but so far isolated, data and code repositories, as well as systems for submission and review management and for publication into a web-based collaborative writing and discussion environment that supports all actors throughout the publication process: readers, authors, and reviewers* (cf. **Figure 1**). The life sciences have furthermore explored the potential of adding services on top of a combination of annotated documents and structured data published on the Web. For example, Saleem et al. (2013) have enabled query answering and visualisation over a large corpus of documents by integrating two, albeit related initially separate datasets, PubMedCentral and The Cancer Genome Atlas. This system shows how high volume and high velocity of latest published bio-medical research papers from PubMed can be intelligently and semantically integrated within the Linked Cancer Genome Atlas dataset to enable new ways of exploration, querying and visualizing big bio-medical data. On a proof-of-concept level the provision of added-value services on top of annotated documents linked to structured data have also been explored in the domain of mathematics (Kohlhase et al. 2011). Exploiting the connection of social sciences articles to data for, e.g., visualisation or data analysis services is not yet in the scope of the OSCOSS project, as we first need to lay the technical foundations for publishing rich annotated and structured documents on the Web. However, previous experience from life sciences and mathematics shows that, in the next step, this foundation enables the provision of added-value services.

## 4. Project Objectives

OSCOSS aims at delivering a new model for scholarly communication in the social sciences – a model that understands the article as an aggregator, a living document that is both an interface to further information on the Web, as well as the pivot for communication and collaboration across scientists. We aim at articles providing the contextual basis on top of which information from other articles, but also from data and code repositories can be reused in a coherent way. This will be enabled by support for adding structured annotations to arbitrarily fine-grained parts of documents, building on interoperable standard models for representing and interlinking knowledge.

---

[11] The German Digital Library (Deutsche Digitale Bibliothek, http://ddb.de) offers open access to Germany's cultural and scientific heritage (8 million books, images, sculptures, music, films etc.).

*End* user groups targeted by our research effort are readers, authors and reviewers of articles in the social sciences, as is shown in the use cases section. However, the responsibility for *enabling* the services for readers, authors and reviewers that have been outlined above is with the *publisher* of social science journals. We thus aim at equipping publishers with a technical platform (cf. **Figure 1**) that puts electronic papers into a central position.

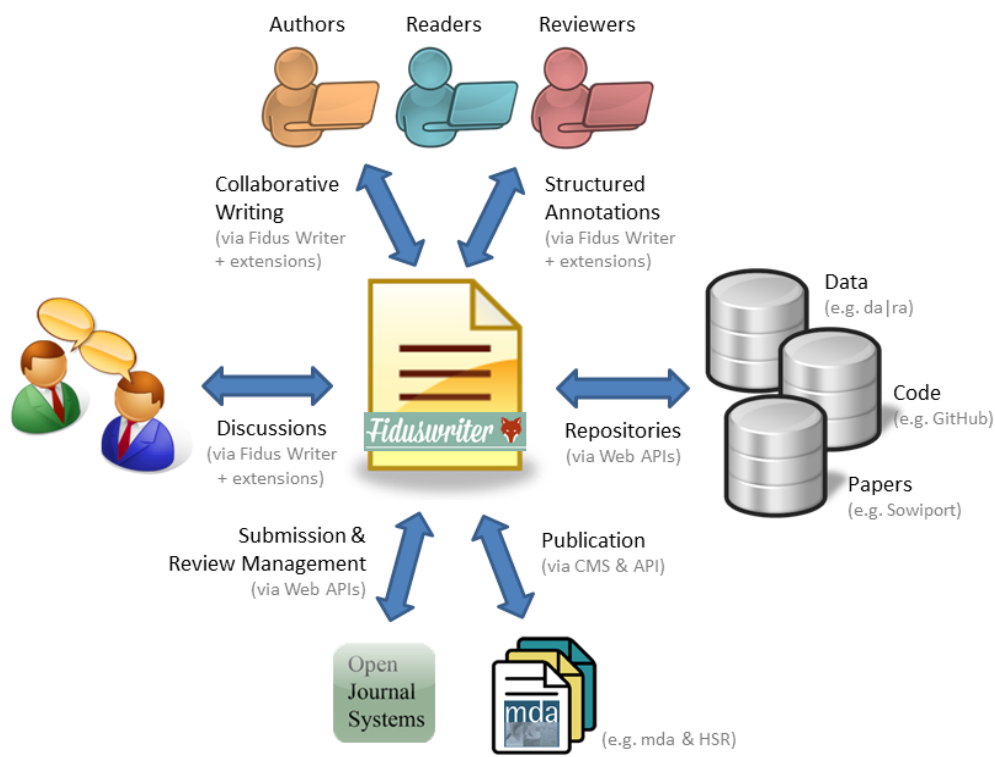

**Figure 1**: System architecture (center: the document; top/left: Fidus Writer extensions; right/bottom: connections to external systems)

**O1. Establishing electronic papers as the focal points of scholarly communication**, i.e. getting rid of media discontinuities throughout the stages of the scientific publication process; in particular:
- making the paper the interactive interface to scientific material (including related articles, as well as data and code), in that the paper provides direct access to the material, e.g., for citation, reuse or inspection,
- providing readers, authors and reviewers with a low-barrier solution for collaborative annotation/bookmarking, writing, knowledge acquisition and reviewing, and with
- a living document as a space for their scholarly communication (e.g., to let authors know exactly what part of a paper or of the data a reviewer's comment applies to, or to enable readers to establish contact precisely with the author who was responsible for the data analysis presented in some table).

*In order to achieve these objectives, it is necessary to …*

**O2. Integrate social science research data, and the papers that interpret them, in a closer way:**
- All material, i.e. articles, data and code, needs to be accessible online.
- Fine-grained parts in articles (e.g. a sentence, a footnote, or a row, column or cell of a table), datasets (e.g. observations of one variable), and source code (e.g. one function) need to be precisely identifiable, so that they can be referenced and annotated.
- Presentations of data in articles, e.g. by tables or figures, should have links to the underlying data or dataset items (for first results, see Ghavimi et al. 2016), and these links should be accessible to readers, authors-as-reusers and reviewers. This enables, with little extra effort, readers to annotate interesting data items, authors to re-analyse the data, and reviewers of an article to inspect the data and assess their quality.

*To ensure acceptance in the social science community, it is necessary to …*

**O3. Keep up the high layout and formatting standards that the community has been used to** from traditional paper-based publication workflows, i.e.,
- to enable citation by exact page numbers and footnote numbers as with paper publications,
- to be compatible with the author guidelines of social science journals (concretely: *mda* and *HSR*[12], both published by GESIS)[13], and
- to render tables and figures in a high visual quality

*Acknowledging that in the scientific community the publisher is the institution that enables these advantages for readers/authors/reviewers, we have to*

**O4. provide publishers with a technical infrastructure satisfying the above objectives, so that they can install and deploy it for their readers/authors/reviewers – including**
- a collaborative environment for writing and discussing articles, connected to data and code repositories, and including an "annotate-only" mode for readers and reviewers,
- a submission and review management system for handling those parts of the submission and review process that are not connected to the document (e.g. submission deadlines, reviewer assignment, editors' decisions), and
- a publishing platform (connecting the reader's view of the collaboration environment, as well as data and code repositories) with interfaces for readers (including authors and reviewers in their "reading role") and for the publisher.

---

[12] Historical Social Research, http://www.gesis.org/hsr/hsr-home/
[13] We chose *mda* and *HSR* because of their significance for social science research and because we have access to the sources of all their publications. We choose two journals instead of one to cover a broader scope of publication practices and workflows; in particular, HSR focuses on the historically-oriented social sciences and mda focuses on empirical and data-driven research. Both journals are heterogeneously structured and well-established major resources which are indexed in the main citation indices (Web of Science and Scopus). Therefore, HSR articles mainly consist of text, footnotes and citations, whereas a larger share of the typical mda article is devoted to tables containing data.

The OSCOSS project addresses the high-level objectives of today's open access transformation technology by developing innovative solutions for

- *both technical and organisational aspects of the publication process*: We intend to design a technical solution that allows to control all steps of the publication process from the electronic paper as a focal point of scholarly communication (see O1 above): from collaborative authoring to submission, to peer review, to online publication, and finally to enriching the reading experience by a facilitated access to data and the possibility to discuss with the authors. While the individual functionalities of our system are derived from the requirements of readers, authors and reviewers, we acknowledge the key role of the publisher as the organisation that provides these services. We therefore aim at developing a bundle of software features for publishers (O4).
- *ensuring the fullest possible reusability of publications*: By embedding structured links to other literature, to data and to code into articles (O2), we aim at increasing the reusability of such artefacts that articles commonly refer to, and at improving the establishment of data and code as (parts of) scientific publications. By enabling the identification and thus citation and annotation of fine-grained parts of articles, and of data and code articles refer to, we aim at facilitating reuse by citation (O2). Despite focusing electronic publication we aim at meeting the high layout and formatting standards of paper-based publishing, so as not to break page-based citation and journal publication workflows (O3).

This effort is primarily driven by GESIS in their role of publishing the mda and HSR social science journals. OSCOSS is innovative because our review of the state of the art (see section "Starting Point") shows that only in the life sciences there are comparable technical solutions combining articles and data. Using such articles throughout all steps of the authoring, reviewing and publishing process is even an entirely new approach. As scholarly communication practice in the social sciences has requirements that have so far been hard to meet with electronic publications, such as citation by page or footnote number, the community would hardly accept an abrupt switch of the mda and HSR publication processes to an electronic, web-based workflow. To minimise the risk of a transition to electronic publishing, we will carefully assess user requirements w.r.t. our objective O3, and carry out usability evaluations w.r.t. our objectives O1–O3.

By choosing two different journals, *mda* and *HSR*, we actually intend to cover a broad scope of scholarly communication practices and publishing workflows, and thus expect our approach to be transferable to other social sciences journals and even to other fields of science. Even more broadly, we aim at developing a generic solution that is ready to be enriched by further functionality and to be applied in other domains as well. As demonstrated by prior research in the life sciences and mathematics, identification and structured annotation of fine-grained parts of articles provide the foundation for making articles self-describing, so further intelligent services can process them.

Looking beyond the duration of the two years of *this* project, we are furthermore laying the foundations for addressing the following objectives:

- *Financing and business models underlying open access*: Making all steps of the publication process accessible from a central collaborative writing environment reduces the costs for copying data about the document into review/submission management systems[14], and the costs for publishing the final article in a separate content management system.

- *Transitioning traditional subscription-based journals to open-access models*: our approach of facilitating online publishing by supporting the authoring of "online-born" articles, and of increasing the reusability of data and code underlying articles by facilitating access to them has the potential to serve as an *incentive* for turning "closed" journals into open access ones.
- *Measuring the impact of open-access publications*: once the published article serves as the central access point to underlying data and code, and once readers can communicate with the authors of an article by using discussion facilities embedded into the published article, we can measure the number of such requests made by readers. Further, structured links to cited literature, to data and to code will enable the definition of metrics that are more fine-grained and thus more precise than impact factors based on the number of citations of an article-as-a-whole.[15]

# 5. Resume

The work done in the OSCOSS project opens scholarly communication in the social sciences and beyond, both in a technical and in a cultural way. On the technical side, we streamline established peer-review workflows by supporting them with web technology, moving from attaching office documents to emails, beyond the current state of the art of isolated web-based word processors, review management systems and data repositories to a coherent collaborative environment for authors, reviewers and, after publication, readers. Once widely in place, beyond the pilots we will run with the mda and HSR journals, this technical solution also has the potential to change the culture of scholarly communication. For example, open, post-publication peer-review can be supported by simply changing a few configuration options in our environment.

# Acknowledgments

The OSCOSS project is funded by the DFG under grant agreements SU 647/19-1 and AU 340/9-1. We thank Fabrizio Orlandi for setting up Figure 1, the System Architecture of the OSCOSS project.

---

[14] A task that is often carried out by the authors and editors, who are not being paid for this, but which is also *supported* by staff of publishing companies

[15] http://altmetrics.org/manifesto/